\documentclass[12pt,english]{paper}
\usepackage[latin9]{inputenc}
\usepackage[letterpaper]{geometry}
\usepackage{color}
\usepackage{babel}
\usepackage{amsmath}
\usepackage{amssymb}
\usepackage{graphicx}
\usepackage{comment}
\usepackage{booktabs}
\usepackage[authoryear]{natbib}
\usepackage{setspace}
\usepackage[unicode=true,pdfusetitle,
 bookmarks=true,bookmarksnumbered=false,bookmarksopen=false,
 breaklinks=true,pdfborder={0 0 0},pdfborderstyle={},backref=false,colorlinks=true]
 {hyperref}
\hypersetup{pdfborderstyle={}, linkcolor=black, urlcolor=blue, citecolor=black}
\usepackage{graphicx}
\usepackage{flafter}
\usepackage{chngcntr}
\setcitestyle{round}

\usepackage{subfig}
\usepackage{xcolor}
\usepackage{hyperref}
\hypersetup{colorlinks, linkcolor={red!50!black},citecolor={blue!50!black},urlcolor={blue!80!black}}

\begin{document}
\vspace{-1.5cm}
\title{Nursing Home Staff Networks and COVID-19\thanks{keith.chen@anderson.ucla.edu, judith.chevalier@yale.edu, elisa.long@anderson.ucla.edu. The authors thank Veraset for access to anonymized smartphone data, and acknowledge research support from the Tobin Center at Yale University. We are grateful for excellent research assistance from Jun Chen, Anna Schickele, and Sabrina Yihua Su, and for the forbearance of Seneca Longchen. This research was found to be not human subjects by the Yale IRB.}}
\author{M. Keith Chen\\
\emph{\small{}UCLA Anderson School of Management}\\
Judith A. Chevalier\\
\emph{\small{}Yale School of Management}\\
Elisa F. Long\emph{\small{}}\\
\emph{\small{}UCLA Anderson School of Management}\\
}
\date{today\\
}
\maketitle
July 22, 2020

\begin{abstract}
\vspace{-0.5cm}
Nursing homes and other long term-care facilities account for a disproportionate share of COVID-19 cases and fatalities worldwide. Outbreaks in U.S. nursing homes have persisted despite nationwide visitor restrictions beginning in mid-March. An early report issued by the Centers for Disease Control and Prevention identified staff members working in multiple nursing homes as a likely source of spread from the Life Care Center in Kirkland, Washington to other skilled nursing facilities. The full extent of staff connections between nursing homes---and the crucial role these connections serve in spreading a highly contagious respiratory infection---is currently unknown given the lack of centralized data on cross-facility nursing home employment. In this paper, we perform the first large-scale analysis of nursing home connections via shared staff using device-level geolocation data from 30 million smartphones, and find that 7 percent of smartphones appearing in a nursing home also appeared in at least one other facility---even after visitor restrictions were imposed. We construct network measures of nursing home connectedness and estimate that nursing homes have, on average, connections with 15 other facilities. Controlling for demographic and other factors, a home's staff-network connections and its centrality within the greater network strongly predict COVID-19 cases. Traditional federal regulatory metrics of nursing home quality are unimportant in predicting outbreaks, consistent with recent research. Results suggest that eliminating staff linkages between nursing homes could reduce COVID-19 infections in nursing homes by 44 percent.

\newpage{}
\end{abstract}

\section{Introduction}

Linked to more than 316,000 COVID-19 cases and 57,000 deaths---nearly half of all U.S. fatalities---nursing homes and other long-term care facilities have been disproportionately afflicted by the ongoing coronavirus pandemic \citep{NYT2020, kaiserdata}. With an elderly resident population, many with underlying chronic medical conditions, congregate living quarters, and routine contact with staff members and outside visitors, nursing homes are particularly vulnerable to outbreaks of respiratory pathogens \citep{strausbaugh2003infectious, lansbury2017influenza}. The U.S. Centers for Medicare and Medicaid Services (CMS), the primary federal regulator of nursing homes, estimates that more than 30 percent of all nursing home residents in New Jersey, Connecticut, and Massachusetts had contracted SARS-CoV-2 as of June 28, 2020 and more than 9 percent of the entire nursing home population died in these states \citep{CMSdata}. 

Evidence from the early outbreak at the Life Care Center of Kirkland, Washington demonstrated that nursing homes and other congregate facilities face extremely elevated risks of virus spread \citep{d2020coronavirus, mcmichael2020covid}. CMS guidance issued on March 13, 2020 significantly restricted visitor access to long-term care facilities---effectively locking down nursing homes to only residents, staff, and contractors \citep{CMSguidance1}. Nevertheless, infections have subsequently broken out in nursing homes, suggesting the unwitting introduction of the virus into homes by staff as one potential channel. In particular, the practice of employing nursing home workers across multiple care facilities may play an important role in the spread of COVID-19, as a U.S. Centers for Disease Control and Prevention (CDC) report issued on March 18, 2020 identified staff working in multiple nursing homes as a likely source of spread from the Life Care Center to other skilled nursing facilities in Washington State \citep{mcmichael2020covid}.

Despite this early recognition of cross-traffic between congregate settings as an important potential transmission mode, the extent of connections between nursing homes remains unknown due to lack of systematic data. Furthermore, although the CDC identified staff members working in multiple long-term care facilities as a key high-risk group, CMS has not provided any specific guidance on this practice nor on reducing contacts between homes more generally \citep{CMSguidance1, CMSguidance2}.

Using novel device-level geolocation data for 509,603 smartphones observed in at least one of the 15,307 nursing homes in the continental U.S., we find that 7 percent of individuals entering a nursing home also entered at least one other nursing home in the six-week period following the March 13th nationwide restriction on nursing home visitors. We construct several measures from network theory to characterize nursing-home connectedness, and examine whether such connectivity predicts confirmed and suspected COVID-19 cases. To our knowledge, this is the first effort to measure and map the network structure of non-social visitors to nursing homes. These data are anonymized, but, given the prohibition of social visitors, this cross-traffic between homes is likely traceable to staff and contractors. We find that the number and strength of connections between nursing homes---and a home's centrality within the greater network---strongly predict COVID cases, even after controlling for location, demographic factors, number of beds, and CMS quality ratings. Consistent with recent research \citep{abrams2020characteristics, Konetzkatestimony, whitecounty}, we observe that traditional federal regulatory metrics of nursing home quality are unimportant in predicting which homes suffered large outbreaks.  

\section{COVID-19 in Congregate Facilities}
The high case count and death toll in long-term care facilities demonstrates the urgent need to understand how transmission mechanisms within these facilities are distinct from broader community spread, to guide targeted policy initiatives and testing strategies \citep{pillemer2020importance, ferguson2020report}.  Given the incomplete case reporting by CMS, extant studies of nursing home cases typically rely upon researcher-compiled state data. Three studies \citep{abrams2020characteristics, Konetzkatestimony, whitecounty} examine the relationship between cases, home location, home demographics, and CMS quality ratings for facilities in a number of states. No study finds CMS ratings to be significant explanators of cases, although demographics and urban location are predictive of cases. Two studies of individual states \citep{he2020there, li2020covid} find that higher CMS-rated nursing homes report fewer cases. While all of these papers provide careful statistical analysis of COVID in nursing home settings, no study directly measures connections amongst homes.  

The importance of connections between congregate settings in SARS-CoV-2 spread has largely been identified through case studies rather than large-scale analysis. The CDC's evaluation of the Kirkland, Washington outbreak pointed specifically to staff employed at multiple nursing homes as a factor in spreading the initial outbreak to additional homes \citep{mcmichael2020covid}. Movement of staff and residents across three affiliated homeless shelters likely contributed to outbreaks in each location \citep{tobolowsky2020covid}. Employees at meat and other food processing plants are at increased risk of contracting SARS-CoV-2 given their proximate working conditions and frequent use of shared transportation between crowded, communal housing and the workplace \citep{dyal2020covid}. 

The movement of incarcerated individuals and the cross-usage of staff across prisons have been identified as risk factors for COVID-19 outbreaks; incoming inmate transfers were the probable source of the San Quentin Prison outbreak \citep{kinner2020prisons, sanquentin}. While we focus on SARS-CoV-2, the importance of linkages between congregate settings has been identified in case studies of prior disease outbreaks. Each of the three flu outbreaks at San Quentin during the 1918 influenza pandemic were linked to the introduction of a single transferred prisoner from a facility where flu was prevalent \citep{sanquentin1918}.

In principle, if a congregate setting were completely closed to the outside, infection could not enter. A key challenge in isolating nursing homes derives from their reliance on staff who live in the community. A study by the State of New York \citep{Newyorkstatedh} concluded, largely based on the timing of infections, that through no fault of their own, nursing home workers were likely the main source of SARS-CoV-2 transmission in nursing homes. They find that roughly one-quarter of nursing home workers in New York State tested positive for the virus. Below, we describe briefly nursing home staffing practices and how they may exacerbate disease spread. 

\section{Nursing Home Staffing Practices and Regulation}
Even in non-pandemic times, nursing home staffing presents challenges. Resident census and health conditions fluctuate from day to day, altering staffing needs on a daily basis with unpredictable absences complicating the staffing problem \citep{slaugh2018consistent}. Understaffing leads to poor service and regulatory violations while overstaffing increases costs. To help manage this trade-off, care facilities often rely on staffing agencies to employ nurses and nurse aides and provide them on an on-call basis \citep{slaugh2018consistent,lu2017mandatory}. While data are limited, a 2009 study suggests that 60 percent of nursing homes use a staffing agency for some of their staffing \citep{castle2009use}. Given this partial reliance on staffing agencies and the recent growth in nursing home chain affiliates \citep{cadigan2015private}, many nurses and nursing assistants commonly work in multiple facilities. Nursing homes also receive services from hospice workers, dialysis technicians, clinicians, medical transporters, and other non-nursing staff that visit multiple homes. In addition to this planned cross-usage, nursing home workers may combine employment across multiple nursing homes as well as other jobs. Survey data from 2012 indicate that 19 percent of nursing assistants and 13 percent of registered nurses hold a second job of some type \citep{secondjob}. According to the Bureau of Labor Statistics, the median nursing assistant earned \$28,980 in May 2019, which makes a willingness to work multiple jobs unsurprising. However, extant regulatory data at the nursing home level do not track the degree to which healthcare workers work in more than one nursing home or other healthcare setting. 

\section{Data and Methodology}

% SUMMARY STATS TABLE
\begin{table}[b!]
\centering 
\caption{Summary statistics of U.S. nursing homes.}
\begin{tabular}{lcc}
\toprule
                                                            & State reporting       & CMS reporting   \\
Variable                                                    & facilities            & facilities \\
\hline
Number of nursing homes                                     & 6,644                 & 12,775   \\
Demographics \\
\quad High proportion ($>$25\%) of Black residents, \%        & 16.1                  & 12.7 \\
\quad High proportion ($>$50\%) on Medicaid, \%               & 33.0                  & 28.3 \\
\quad Urban location, \%                                    & 80.5                  & 72.5 \\
Regulatory measures \\
\quad Number of beds                                        & 114 (58.5)            & 110 (60.6) \\
\quad CMS quality rating (1-5)                              & 3.18 (1.42)           & 3.15 (1.42) \\
\quad Has infection violations, \%                          & 75.3                  & 75.7 \\
Network metrics \\
\quad Node degree                                           & 15.6 (17.5)   & 14.3 (16.5) \\
\quad Node strength                                         & 20.9 (26.3)   & 19.7 (30.8) \\
\quad Weighted average neighbor degree                      & 24.4 (20.7)   & 24.5 (33.6) \\
\quad Eigenvector centrality in state                       & 0.14 (0.22)   & 0.13 (0.22) \\
\hline
\end{tabular}
\vspace{-0.5cm}
\begin{flushleft}
\vspace{0.2in}
\small {CMS facilities include all continental U.S. nursing homes that report demographic and regulatory data. Binary variables are \% of nursing homes; continuous variables are mean values with standard deviations in parentheses.}
\end{flushleft}
\label{tab:summarytable}
\end{table}

Examination of the nursing home COVID-19 crisis is hindered by the fact that CMS did not require nursing homes to submit data on COVID-19 cases and fatalities until May 2020. For our main data analysis, we use the disclosures of individual state Departments of Public Health to determine cumulative nursing home COVID cases. From the 23 states for which home-level resident case data are available, we collected data on cumulative resident cases as of May 31, 2020 (or closest reporting period). In the Supplementary Information, we repeat our analyses using the cumulative case data reported by CMS for homes nationwide, with the caveat that CMS instructions for reporting cumulative cases allowed nursing homes to not report cases occurring before May 2020. 

Using the CMS address of record for each facility, we merge the nursing home-level COVID-19 case data with nursing home staff-network connections measured using anonymized device-level smartphone data for the continental U.S. over the period March 13 to April 23, 2020. CMS had announced guidelines preventing nursing home social visitors starting March 13, 2020.  Thus, over the period for which smartphone activity was measured, the smartphones detected in nursing homes would likely belong to staff, contractors, and residents. Summary statistics of nursing homes for both the 23 states for which we have assembled state data and the set of facilities regulated by CMS in the continental U.S. for which we have complete data are found in Table \ref{tab:summarytable}.

\subsection*{Smartphone Location Data}
We estimate staff networks across nursing homes using anonymized smartphone-location data provided by Veraset, a company that aggregates location data across several apps on both the Apple and Android platforms after the smartphone user has consented to the use of their anonymized data. Previous studies of these data have found them to be highly representative of the U.S. on numerous demographic dimensions \citep{chen2018thanksgiving}. Of the roughly 30 million smartphones we include in our U.S. sample, we identify 509,603 smartphones that appear to enter at least one U.S. nursing home in the six-week period of our study.

We match all U.S. nursing homes with a shapefile delineating each facility's rooftop boundary. To do so, we match a nursing home's CMS-provided street address to a latitude-longitude location through the use of the Google Maps API, and then match that location to a satellite-image machine-learned shapefile of the convex-hull of the building's rooftop (provided by Microsoft / Open Streetmaps). Using these rooftop shapefiles, we find all smartphones in our data that have ever recorded more than two location traces in the same nursing home facility during our study period, when visitors were explicitly barred. By identifying smartphones that entered more than one nursing home, we estimate a nursing home staff-contact network.

\subsection*{Network Metrics}
The contact structure among nursing home facilities within a state is represented by an undirected network consisting of $n$ nodes (the facilities) and $n(n-1)/2$ possible edges (pairs of facilities). We estimate a symmetric $n \times n$ adjacency matrix $\mathbf{A}$, where $a_{ij}=1$ if at least one smartphone is observed in both facilities $i$ and $j$, and 0 otherwise. Edge weights $w_{ij}$ correspond to the number of smartphones observed in both facilities.   

A facility's \textit{degree} $k_i$ equals the total number of other nursing home facilities with a connection to facility $i$ (i.e., the number of \textit{neighbors} of node $i$). 
\begin{equation}
k_i = \sum_{j=1}^n a_{ij}
\end{equation}

\textit{Strength} $s_i$ is the weighted sum of connections to other facilities (i.e., the total number of smartphones that appear in facility $i$ and some other nursing home). 
\begin{equation}
s_i = \sum_{j=1}^n w_{ij} a_{ij}
\end{equation}

\textit{Weighted average neighbor degree} $\bar{k}^w_i$ is the average degree of node $i$'s neighbors (i.e., the neighbors' connections to other facilities), weighted by the number of connections $w_{ij}$ shared with node $i$, as previously defined \citep{barrat2004architecture}. 
\begin{equation}
\bar{k}^w_i = \frac{1}{s_i}\sum_{j=1}^n w_{ij} a_{ij} k_j
\end{equation}

\textit{Eigenvector centrality} $v_i$ measures the extent to which node $i$ is connected to other highly connected nodes in the network. 
\begin{equation}
v_i = \frac{1}{\lambda}\sum_{j=1}^n a_{ij} v_j
\end{equation}
This centrality measure is computed using the principal eigenvector of the adjacency matrix, rewritten in matrix notation as $\mathbf{Av}=\lambda\mathbf{v}$. We then normalize these values to range between 0 and 1 within each state's network.

\subsection*{Empirical Specification}
Our main specification examines the predictors of the number of cases in a nursing home as a function of various explanatory variables. Specifically, we include the home's \emph{demographic} characteristics, including the number of beds and the squared number of beds to allow for nonlinearities. Following previous literature, we include indicator variables for whether a nursing home has a large proportion ($>50\%$) of residents on Medicaid and a large proportion ($>25\%$) of Black residents. We also include \emph{CMS quality} measures of nursing homes as done previously \citep{abrams2020characteristics}. CMS rates nursing homes on a five-point scale. Our specification includes indicator variables for one-star, two-stars, three-stars, and four-stars, with the omitted category being five-stars (the highest possible rating). We include a binary indicator variable that equals one if the home had \emph{infection control violations} in its most recent inspection. We define an indicator variable for whether a home is in an \emph{urban} location based on the CDC's urban-rural classification \citep{CDC2020}.

We examine whether nursing home connectivity predicts COVID-19 cases using the following regression model:
\begin{align*}
sinh^{-1}(Cases_i) &= \boldsymbol{\beta}_0 + \boldsymbol{\beta}_1 NodeDegree_i + \boldsymbol{\beta}_2 NodeStrength_i\\
          & + \boldsymbol{\beta}_3 WeightNeighDeg_i + \boldsymbol{\beta}_4 EigenCentrality_i\\
          &+ \boldsymbol{\gamma}_0 \mathbf{X}_i + \boldsymbol{\gamma}_1 \mathbf{F}_i + \varepsilon_i
\end{align*}
where $sinh^{-1}(x) = \ln(x+\sqrt{1+x^2})$ is the inverse hyperbolic-sine of a nursing home's COVID-19 cases. All reported semi-elasticities are adjusted for the $sinh^{-1}$ functional form. We include as independent variables the four network measures that characterize a facility's network connectivity, as described in the previous section. The vector $\mathbf{F}_i$ is a set of fixed effects, for the geographical unit of the home (state or county), and the vector $\mathbf{X}_i$ includes demographic, geographic, and regulatory controls for nursing home $i$.

To control for reporting and other differences across states, we include state fixed effects. In the Supplement, we include coefficients for all variables in Table \ref{tab:resultstabledetailed}. Table \ref{tab:resultstablebinary} replaces the dependent variable with a binary indicator variable if the nursing home has had any COVID-19 cases. Table \ref{tab:resultscountyfe} repeats our analysis using county fixed effects. This robustness check is important given the finding in \citep{whitecounty} that county-level SARS-CoV-2 prevalence predicts case counts in nursing facilities. Finally, Table \ref{tab:resultscms} replaces the data for 23 states with the larger CMS dataset for the continental U.S.

\section{Network Measures}
\begin{figure}[b!]
\caption{Degree distribution of nursing homes with and without COVID cases (reported to CMS as of May 31, 2020).}
\centering
\includegraphics[width=0.6\linewidth]{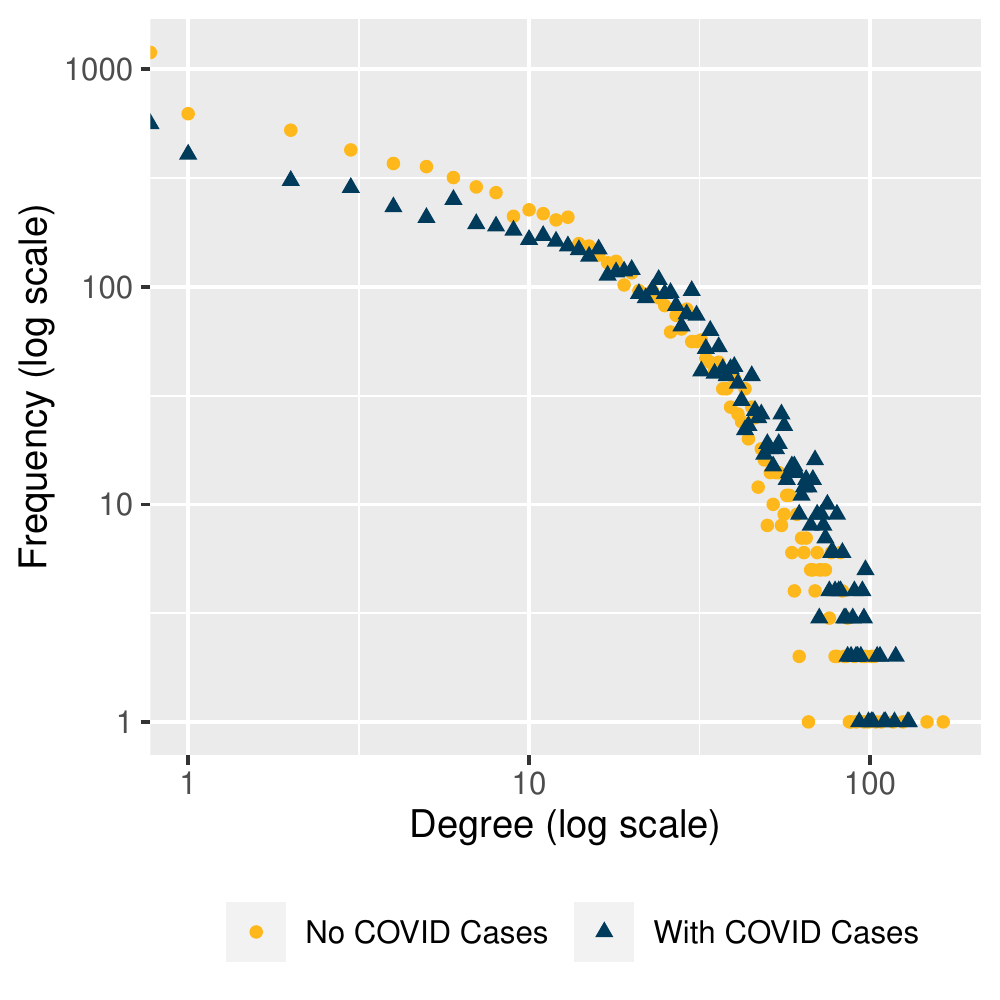}
\label{fig:degreedistribution}
\end{figure}

Nursing homes display a wide range of connectedness to other facilities. Average \textit{degree}---the number of other facilities a nursing home shares at least one smartphone connection with---across all U.S. nursing homes is $\langle k \rangle=14.3$, but this ranges from a state-average degree below 2 in Montana, South Dakota, Vermont, and Wyoming, to an average exceeding 20 in Florida, Illinois, Maryland, New Jersey, and Texas. Among nursing homes with confirmed or suspected cases, as reported to CMS, average degree is $17.3$ compared to $11.8$ among homes with no documented cases ($t=19.2$, $p<0.0001$), with a similar effect across the entire degree distribution (see Fig. \ref{fig:degreedistribution}). Average \textit{strength}---the total number of smartphones appearing in a nursing home and one of its neighbors---is also greater in COVID-positive homes (23.7 vs. 16.4, $t=13.4$, $p<0.0001$).

% DEGREE DISTRIBUTION

\begin{figure*}[tbp]
\caption{Network structure of selected nursing home facilities in six U.S. states. Details provided in Table \ref{tab:networktable}. \\[-4ex]} 
\centering
\subfloat{\includegraphics[width=0.45\textwidth]{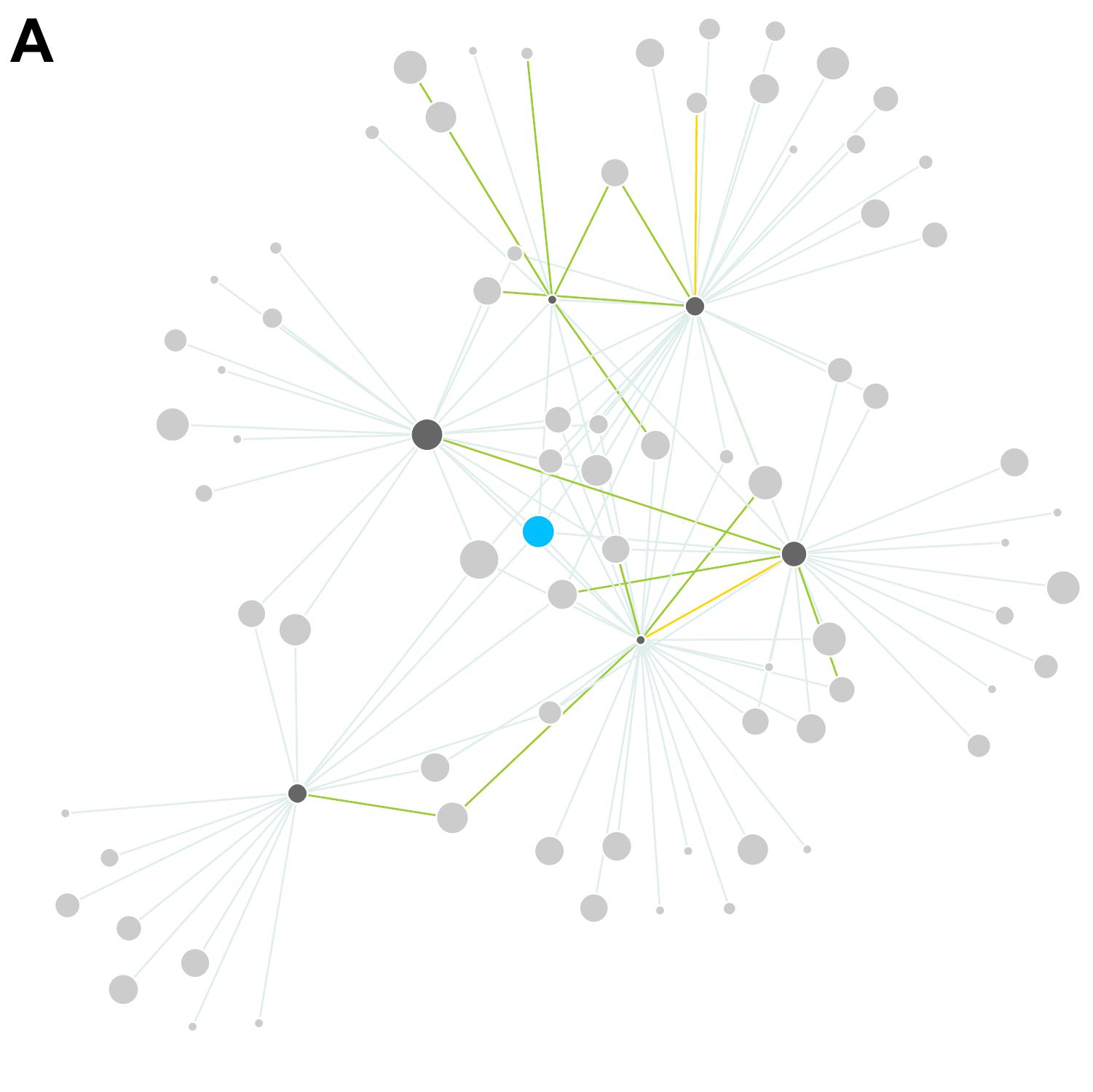}}
\subfloat{\includegraphics[width=0.45\textwidth]{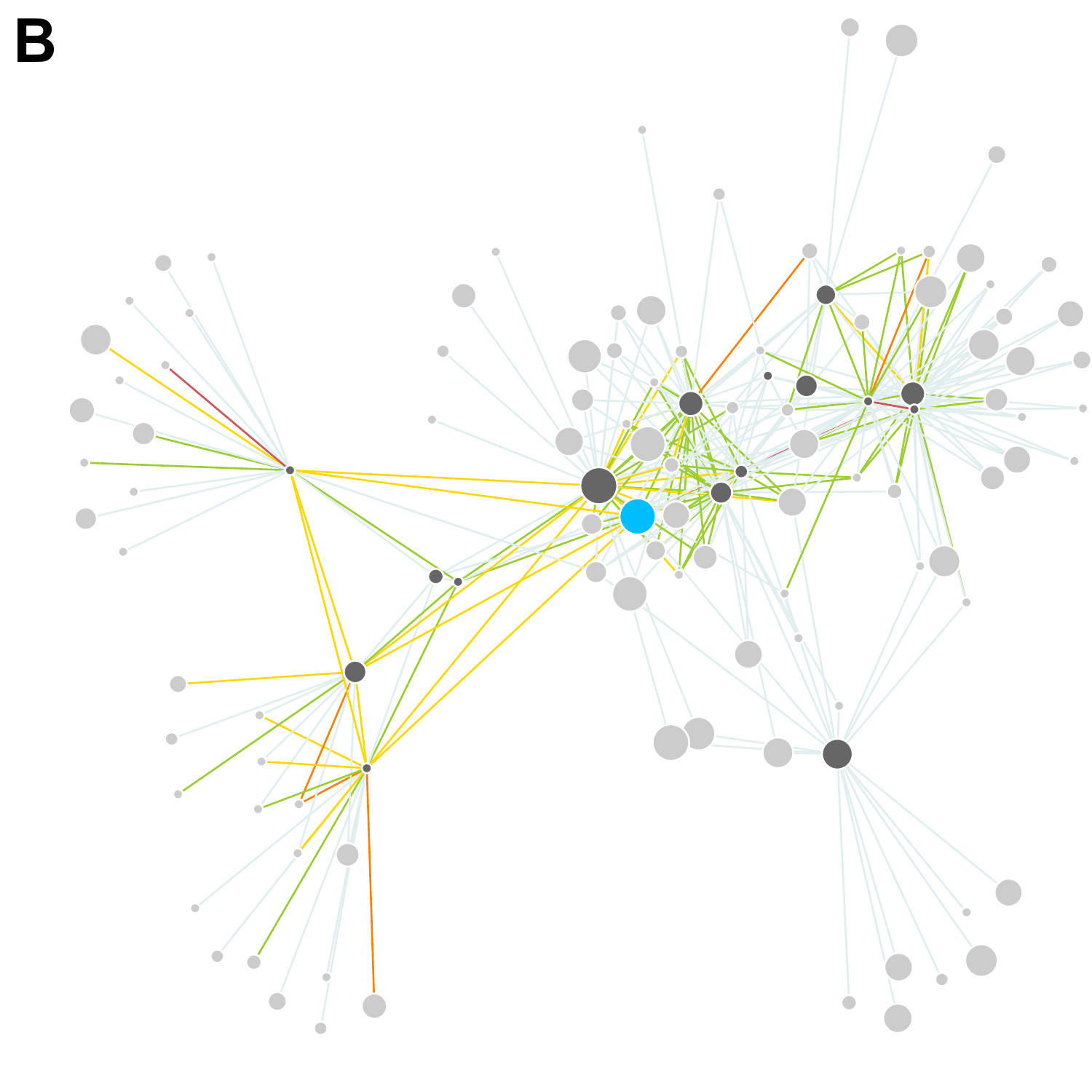}}
\\[-4ex]
\subfloat{\includegraphics[width=0.45\textwidth]{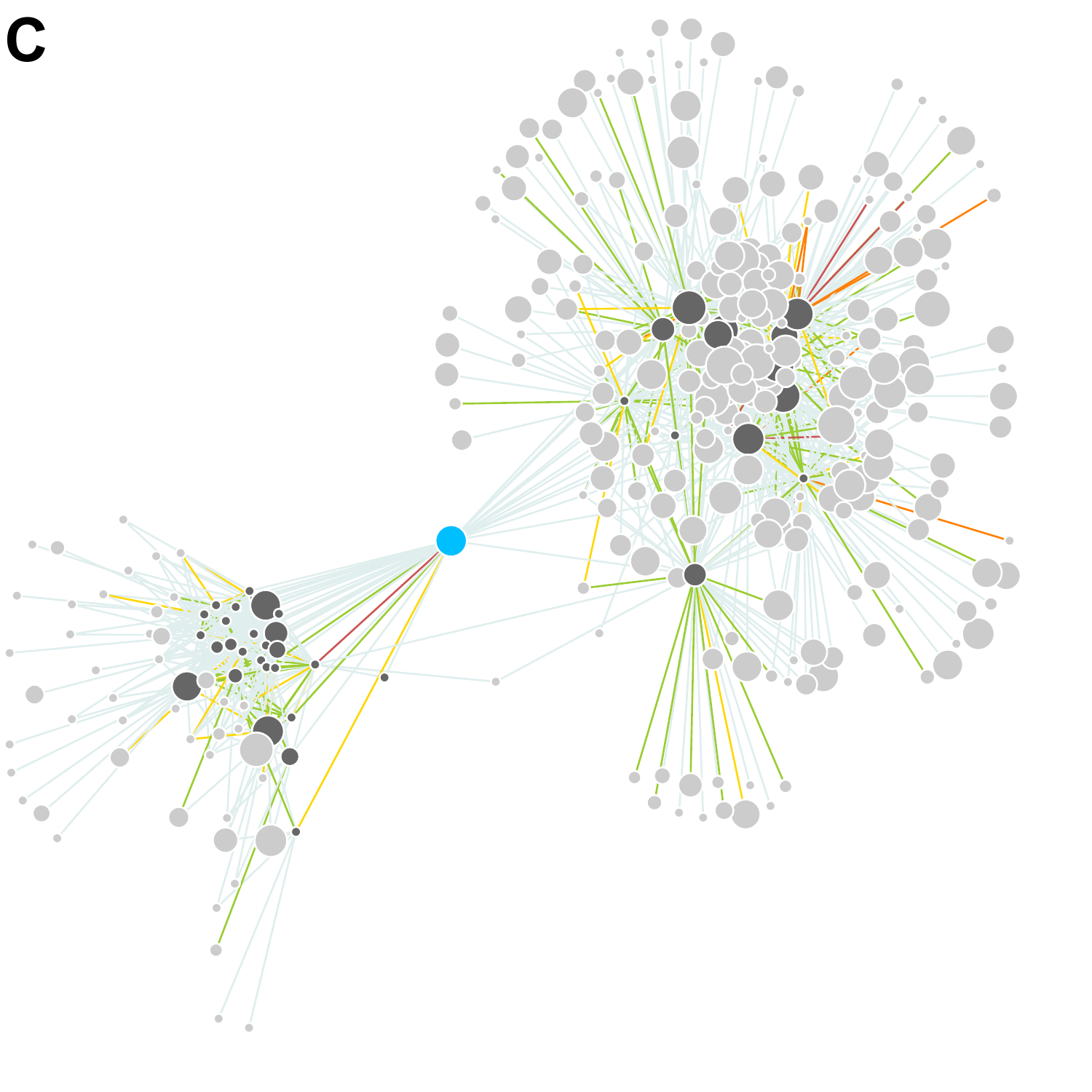}} 
\subfloat{\includegraphics[width=0.45\textwidth]{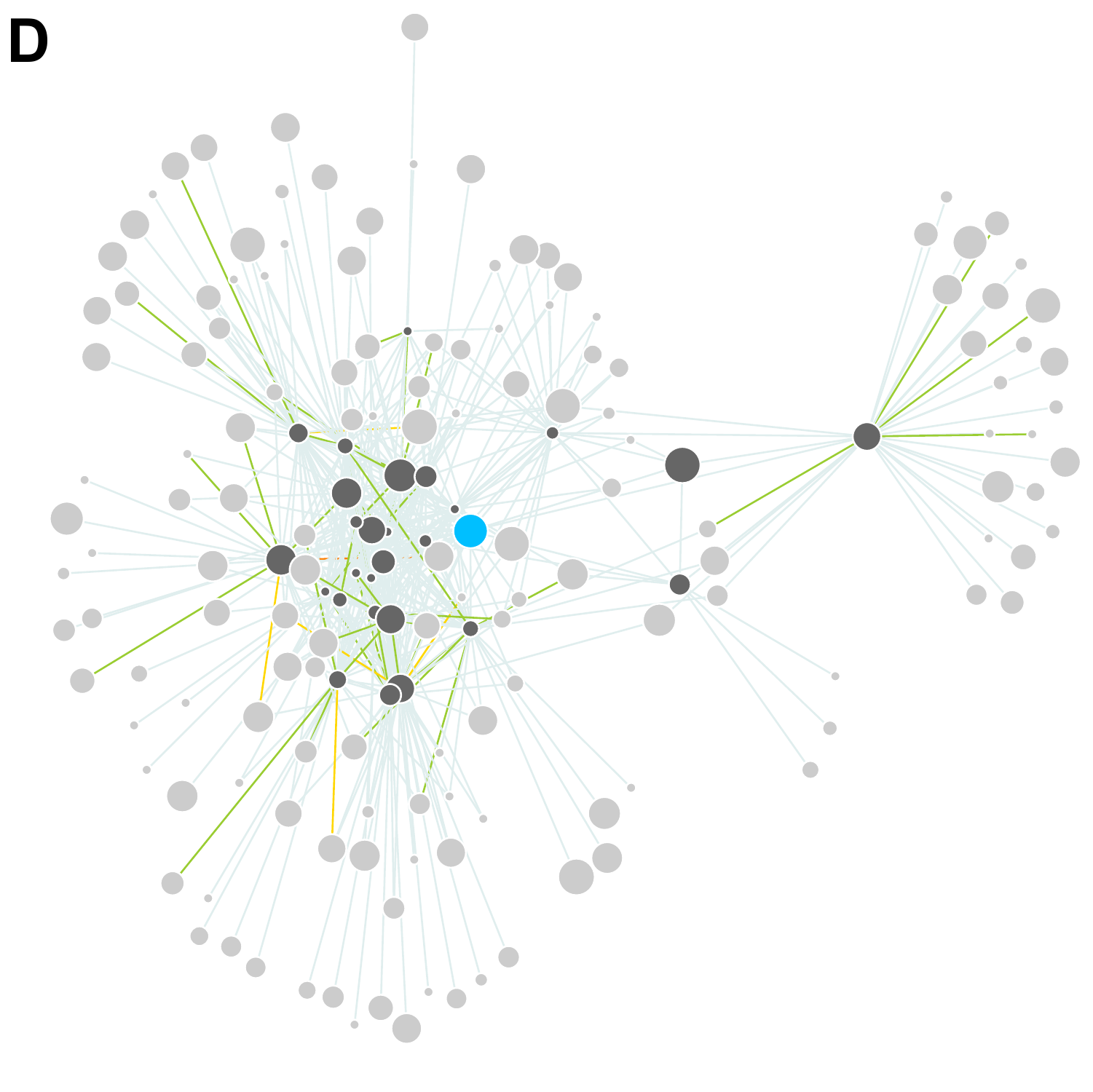}}
\\[-4ex]
\subfloat{\includegraphics[width=0.45\textwidth]{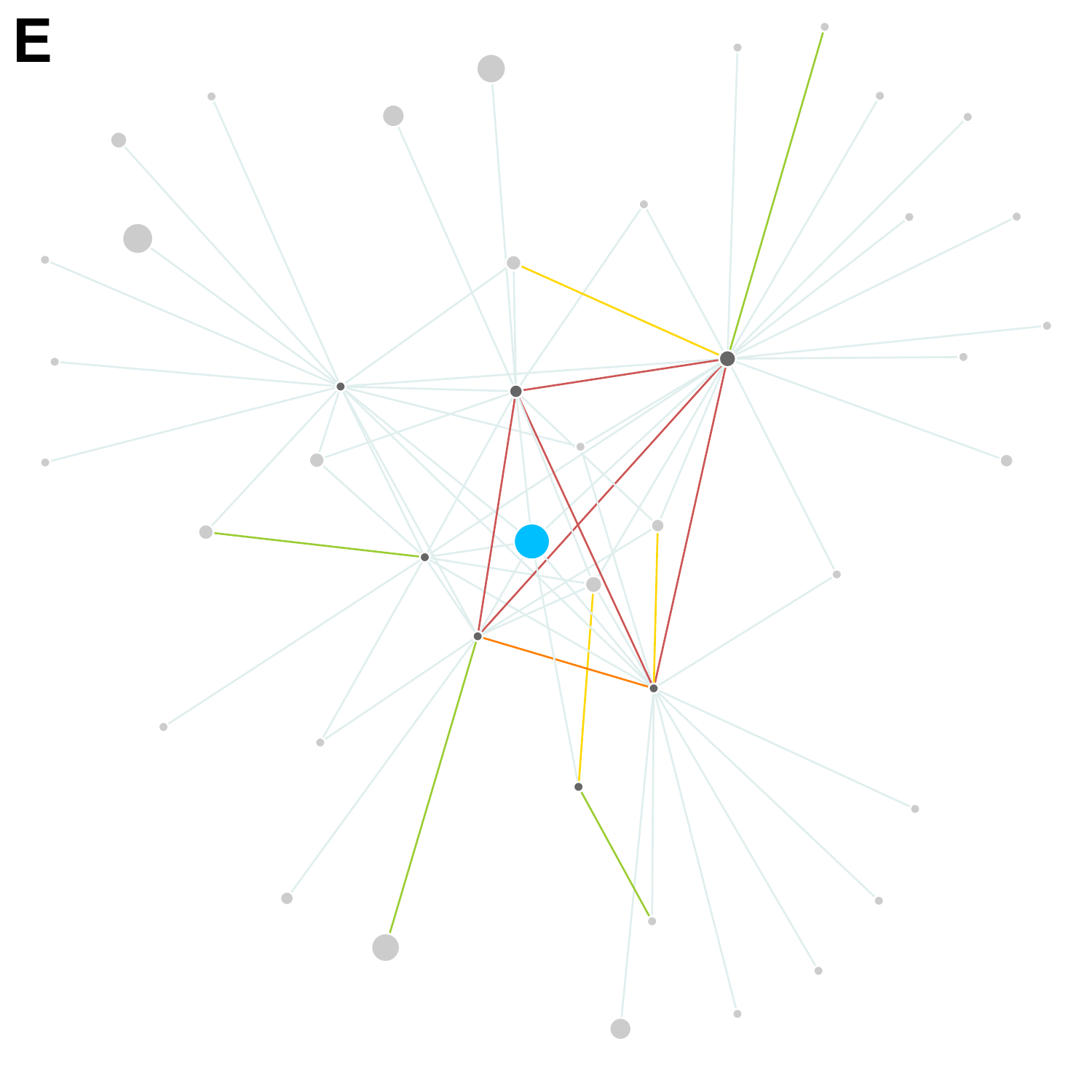}} 
\subfloat{\includegraphics[width=0.5\textwidth]{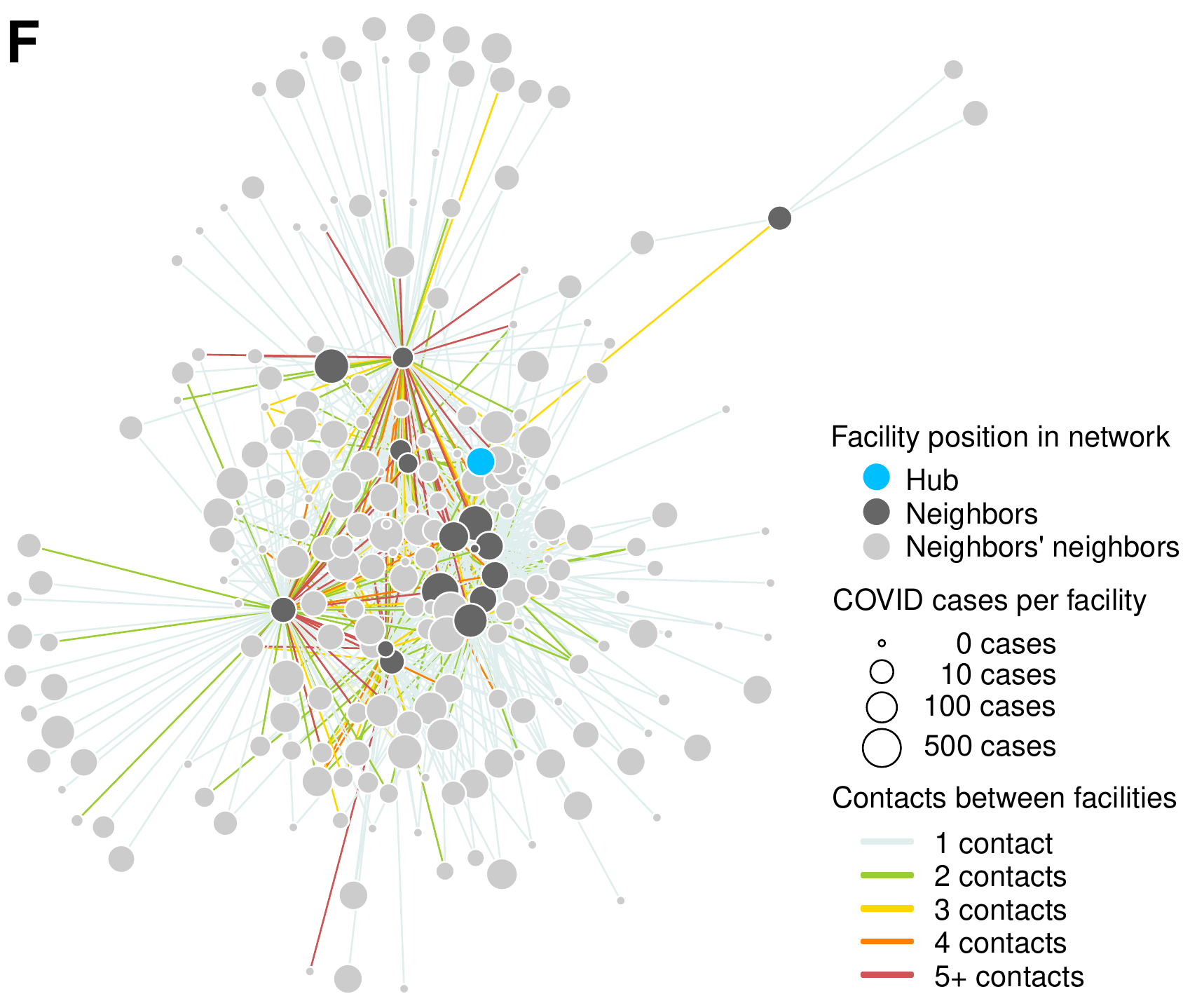}} 
\label{fig:networkfigure}
\end{figure*}

To illustrate how degree, strength, and other network measures differ across nursing homes, we present  network diagrams for a subset of homes in six states as depicted in Fig. \ref{fig:networkfigure} and summarized in Table \ref{tab:networktable}. Here, nodes represent individual nursing homes and edges represent connections between nodes (i.e., at least one smartphone observed in both homes). More connected nodes are generally towards the center of each diagram and nodes with fewer connections are on the periphery. In each sub-network, a focal nursing home or ``hub'' is shown in blue, with its direct neighbors (homes with at least one contact with the hub) in dark grey and its neighbors' neighbors in light grey. Node size corresponds to the number of confirmed and suspected COVID-19 cases (from CMS data to allow for continental U.S. coverage) and edge color corresponds to the number of smartphones observed in that pair of homes.

% NETWORK FIGURE

A major challenge facing nursing homes is that every connection is a potential link to other connections---and to SARS-CoV-2 transmission. In the Connecticut sub-network (Fig. \ref{fig:networkfigure}A), for instance, the focal nursing home had a large outbreak of 102 cases, yet is directly connected to only six other homes (\emph{degree} = 6), with one smartphone observed in each pair (\emph{strength} = 6), well below our sample mean or Connecticut's average degree of 13. However, the hub's neighbors are highly connected themselves, resulting in a \emph{weighted average neighbor degree} of 26, well above the state's average. 

% NETWORK CENTRALITY TABLE
\begin{table}[t!]
\centering 
\caption{Network centrality measures for six selected nursing homes.}
\begin{tabular}{ccccccc}
\toprule
Hub      &          & COVID-19  &         &           & Wtd. Avg.   & Eigenvector \\ 
Facility & State    & Cases     & Degree  & Strength  & Neigh. Deg. & Centrality  \\
\hline
A        & CT       & 102       & 6       & 6         & 26.2        & 0.13 \\
B        & GA       & 189       & 16      & 30        & 18.2        & 0.15 \\
C        & IL       & 72        & 42      & 51        & 45.1        & 0.12 \\
D        & MA       & 134       & 27      & 27        & 32.5        & 0.75 \\
E        & NC       & 181       & 7       & 7         & 23.0        & 0.02 \\
F        & NY       & 57        & 16      & 36        & 61.1        & 0.17 \\ 
\bottomrule
\end{tabular}
\label{tab:networktable}
\end{table}

The hub node in Georgia (Fig. \ref{fig:networkfigure}B) has average degree, but with strength double the state's average, an indicator that many staff members are likely working at this facility and its 16 neighbors. Crucially, these individuals are potential conduits of SARS-CoV-2 transmission given the home's 189 cases by the end of May, one of the largest outbreaks in the state. Likewise, the hub node in the Illinois sub-network (Fig. \ref{fig:networkfigure}C) is not the most connected home in the state with a \emph{degree} of 42, yet this facility can play an important role in viral spread by acting as a bridge between geographically distant nursing homes. In Massachusetts (Fig. \ref{fig:networkfigure}D), the selected hub node is quite central to this network, with an eigenvector centrality of 0.75, substantially higher than the state average of 0.16, because this particular facility is directly connected to many other central nodes. The North Carolina network (Fig. \ref{fig:networkfigure}E) illustrates a hub node connected to each of four other nursing homes, which share several contacts among themselves. Here, the focal facility faced a large outbreak of 181 cases, potentially increasing the risk of transmission to its neighboring homes. Lastly, with more than 600 nursing homes in New York State (Fig. \ref{fig:networkfigure}F), average degree is higher than in most states. With a weighted average neighbor degree of 61, this particular hub node is connected to several other central nodes, but unlike the Massachusetts network, many nursing home pairs in New York appear to share multiple contacts.

\section{Regression Results}
Table \ref{tab:resultstable} presents multivariate regressions of cumulative nursing home COVID-19 cases as of May 31 on a set of explanatory variables. Importantly, these regression specifications include state fixed effects to allow for differences in baseline risks and reporting practices across states; we include even finer county fixed effects in Supplementary Table S3. We use the inverse hyperbolic-sine of cases as the dependent variable, given its non-negative skewed distribution. Column (1) shows our base specification with our simplest network explanatory variable, \emph{node degree} $k_i$---the number of ``neighbors'' or other nursing homes connected to the focal home by at least one smartphone. Results indicate that, if a home adds 10 neighbors (average degree is 15), the expected number of COVID-19 cases increases by $13.7 \times 1.95=26.7\%$. Column (2) replaces the degree measure with \emph{node strength} $s_i$---the total number of ``contacts'' or smartphones that appear in the facility of interest and in some other nursing homes. This too predicts nursing home cases significantly: if a home adds 10 contacts (average strength is 21), expected cases of COVID-19 increases by $6.79 \times 1.78=12.1\%$. Column (3) replaces the degree and strength measures with \emph{weighted average neighbor degree} $\bar{k}^w_i$---the weighted average degree of a nursing home's neighbors \citep{barrat2004architecture}. Here, an increase of 10 (mean is 25.6) would lead to an expected $14.7 \times 2.09=30.7$ percent increase in cases. Column (4) uses our final network measure, \emph{eigenvector centrality} $v_i$---which measures the extent to which the nursing home is connected to other highly connected nursing homes, normalized to range between 0 and 1 within each state. This measure implies that, as we move from an unconnected nursing home in the state ($v_i=0$) to the most connected ($v_i=1$), the total increase in expected cases exceeds 190\%.

Column (5) is our final specification, which includes both weighted average neighbor degree and eigenvector centrality, summarizing both the local and global connectivity of a nursing home. Intuitively, regression 5 compares demographically and geographically situated nursing homes of similar quality, which are thus likely exposed to similar risk of community-spread. When infections co-vary with differences in staff-network connectivity and centrality, these increased infections are likely attributable to shared-staff transmitting the virus across multiple nursing homes. As a useful counterfactual, regression 5 suggests that severing all staff-linkage transmission between facilities could reduce nursing home resident COVID-19 cases by 43.6\%.

% REGRESSION RESULTS TABLE
\begin{table}[b!]
\centering 
\caption{Covariates of COVID-19 cases within nursing homes.}
\begin{tabular}{lccccc}
\toprule
& \multicolumn{5}{c}{Dependent variable: $sinh^{-1}(Cases)$} \\
\cmidrule(lr){2-6}
                                                & (1)       & (2)           & (3)       & (4)       & (5) \\ 
\midrule
{Urban indicator}                          & 0.649***  & 0.686***      & 0.609***  & 0.650***  & 0.574*** \\ 
	                                            & (0.054)   & (0.054)       & (0.053)   & (0.053)   & (0.054) \\ 
{Node degree} \rule{0pt}{3ex}              & 0.0137*** &               &           &           &  \\ 
	                                            & (0.0012)  &               &           &           &  \\
{Node strength} \rule{0pt}{3ex}            &           & 0.00679***    &           &           &  \\ 
	                                            &           & (0.00079)     &           &           &  \\ 
{Weighted average} \rule{0pt}{3ex}         &           &               & 0.0147*** &           & 0.0116*** \\ 
{neighbor degree}	                        &           &               & (0.0010)  &           & (0.0011) \\ 
{Eigenvector centrality} \rule{0pt}{3ex}   &           &               &           & 1.073***  & 0.643*** \\ 
{in state}	                                &           &               &           & (0.091)   & (0.099) \\ 
{Fixed effects}  \rule{0pt}{3ex}           & State     & State         & State     & State     & State \\ 
{Home demographics} \rule{0pt}{3ex}        & Yes       & Yes           & Yes       & Yes       & Yes \\ 
{CMS quality rating} \rule{0pt}{3ex}       & Yes       & Yes           & Yes       & Yes       & Yes \\ 
\midrule
{Observations}                             & 6,644     & 6,644         & 6,644     & 6,644     & 6,644 \\ 
$F$-stat                                        & 123.8     & 117.9         & 132.0     & 125.0     & 125.3 \\ 
$R^2$                                           & 0.412     & 0.407         & 0.419     & 0.413     & 0.422 \\ 
{Within} $R^2$                             & 0.169     & 0.163         & 0.179     & 0.171     & 0.184 \\
\bottomrule
\end{tabular}
\vspace{0.1cm}
\begin{flushleft}
\footnotesize {Standard errors in parentheses. Significance levels: $^+ p<0.05$, $^* p<0.01$, $^{**} p<0.001$, $^{***} p<0.0001$. \\
Dependent variable is inverse hyperbolic sine of COVID cases using state data. Demographics include number of beds, high proportion of Black residents, and high proportion on Medicaid. CMS quality is a 1-5 categorical rating.}
\end{flushleft}
\label{tab:resultstable}
%\vspace{-0.5cm}
\end{table}

Consistent with other studies \citep{abrams2020characteristics,Konetzkatestimony, whitecounty}, we find that CMS ratings of nursing home quality are not predictive of infections, yet facilities in urban locations, those with more beds, a higher share of Black residents, or a higher share of residents on Medicaid are all associated with more COVID-19 cases (details in Supplementary Table \ref{tab:resultstabledetailed}).

\section{Discussion and Conclusions}
Using a large-scale analysis of smartphone location data, we document substantial connections amongst nursing homes after nationwide visitor restrictions were enacted in March 2020. Consistent with the CDC's conclusion that shared workers were a source of infection for the nursing home outbreak in Kirkland, Washington \citep{mcmichael2020covid}, our network measures suggest that individuals moving between nursing homes is a significant predictor of SARS-CoV-2 infections. Our general findings are robust to alternative specifications or the use of the case count data available from CMS. Clearly, there are limitations to any observational study. However, this is not an environment in which randomized controlled trials are feasible or ethical. 

These results provide evidence of the magnitude of the benefits that would derive from compensating nursing home workers to work at only one home and limiting cross-traffic across homes. While some nursing homes and other long-term care facilities have undertaken actions to create a ``staff bubble'', this is still not a component of extant regulation \citep{bubble, baltimorenursing}. Absent such regulation, allocation of PPE, testing, and other preventative measures should be targeted thoughtfully, recognizing the current potential for cross-transmission across homes. While the nursing home population is particularly fragile, this research has implications for cross-linkages amongst other congregate settings such as assisted living homes, prisons, or large workplace facilities such as food-processing plants.

\clearpage
\section{Bibliography}

\renewcommand{\bibsection}{}
\bibliographystyle{abbrvnat}

{\raggedright
\bibliography{papers}
}

%%\phantomsection\addcontentsline{toc}{section}{\refname}\bibliography{../Dropbox/Paper/papers,papers}
%\appendix

\clearpage
\section{Supplementary Materials}
\counterwithin{table}{section}
\setcounter{table}{0}
\renewcommand\thetable{S\arabic{table}}

In this Supplement, we provide tables with additional detail from our Table \ref{tab:resultstable} and three robustness checks of our main empirical specification. First, we show all coefficient estimates for our main analysis (Table \ref{tab:resultstabledetailed}). Second, we repeat our main analysis, replacing the inverse hyperbolic sine of the number of cases in the nursing home with a binary variable that takes the value of 1 if the nursing home has cases (Table \ref{tab:resultstablebinary}). Here, for example, the results in column (1) suggest that, when 10 additional nursing home connections are added, the probability that a home has cases increases by 3 percentage points. In the data overall, 42 percent of homes have cases. The network measures are all statistically significant in this alternative specification. Third, we repeat our main analysis replacing the state fixed effects with county fixed effects (Table \ref{tab:resultscountyfe}). This allows a smaller number of units within which variation can be measured. The urban variable, which is measured at the county-level, is omitted. Results are qualitatively similar to our state fixed effects data though, as expected, significance levels diminish somewhat.

As a final robustness check, we repeat our main analysis using data from CMS rather than data from the individual states (Table \ref{tab:resultscms}). This allows us to examine the 48 continental United States plus the District of Columbia, but these data are subject to the reporting limitation that homes were not required to add cases prior to May into their cumulative case totals. The CMS data reports cases in 45 percent of nursing homes but reports overall fewer cases than the individual state data. This is expected since the CMS data did not require homes to report cases in the cumulative total that had resolved before May 2020. In this robustness specification, point estimates for the network variables are slightly smaller than in our base specifications but qualitatively extremely similar. 

Finally, we report mean values and standard deviations of our four network measures, for all 48 contiguous U.S. and the District of Columbia (Table \ref{tab:statenetwork}).

% TABLE S1 - DETAILED REGRESSION RESULTS 
\begin{table}
\centering \small 
\caption{Detailed covariates of COVID-19 cases within nursing homes.}
\begin{tabular}{lccccc}
\toprule
& \multicolumn{5}{c}{Dependent variable: $sinh^{-1}(Cases)$} \\
\cmidrule(lr){2-6}
                                                & (1)           & (2)           & (3)           & (4)           & (5) \\ 
\midrule
{Beds}                                     & 0.00920***    & 0.00974***    & 0.00983***    & 0.00948***    & 0.00933*** \\ 
	                                            & (0.000721)    & (0.000721)    & (0.000709)    & (0.000717)    & (0.000711) \\ 
{Beds}$^2$ \rule{0pt}{3ex}                 & -0.00000496*  & -0.00000566** & -0.00000527*  & -0.00000517*  & -0.00000475* \\ 
	                                            & (0.00000167)  & (0.00000167)  & (0.00000165)  & (0.00000166)  & (0.00000165) \\ 
{High proportion} \rule{0pt}{3ex}          & 0.105$^+$     & 0.112*        & 0.0833        & 0.112*        & 0.0888$^+$ \\
{on Medicaid}	                            & (0.0433)      & (0.0434)      & (0.0431)      & (0.0432)      & (0.0430) \\ 
{High proportion} \rule{0pt}{3ex}          & 0.533***      & 0.551***      & 0.519***      & 0.515***      & 0.485*** \\ 
{of Black residents}	                    & (0.0575)      & (0.0577)      & (0.0571)      & (0.0576)      & (0.0572) \\ 
{CMS rating 1} \rule{0pt}{3ex}             & 0.0292        & 0.0359        & 0.00298       & 0.0253        & 0.0105 \\ 
	                                            & (0.0660)      & (0.0663)      & (0.0657)      & (0.0660)      & (0.0655) \\ 
{CMS rating 2} \rule{0pt}{3ex}             & 0.0587        & 0.0681        & 0.0402        & 0.0573        & 0.0436 \\ 
	                                            & (0.0602)      & (0.0604)      & (0.0599)      & (0.0601)      & (0.0597) \\ 
{CMS rating 3} \rule{0pt}{3ex}             & 0.110         & 0.119$^+$     & 0.0935        & 0.114         & 0.0982 \\ 
	                                            & (0.0604)      & (0.0606)      & (0.0600)      & (0.0603)      & (0.0598) \\ 
{CMS rating 4} \rule{0pt}{3ex}             & 0.0132        & 0.0210        & 0.00337       & 0.0184        & 0.0100 \\ 
	                                            & (0.0560)      & (0.0562)      & (0.0557)      & (0.0559)      & (0.0555) \\ 

{Has infection}\rule{0pt}{3ex}     & -0.0709       & -0.0676       & -0.0511       & -0.0681       & -0.0573 \\ 
{violations}          & (0.0485)      & (0.0487)      & (0.0482)      & (0.0484)      & (0.0481) \\ 
{Urban indicator} \rule{0pt}{3ex}          & 0.649***      & 0.686***      & 0.609***      & 0.650***      & 0.574*** \\ 
	                                            & (0.054)       & (0.054)       & (0.053)       & (0.053)       & (0.054) \\ 
{Node degree} \rule{0pt}{3ex}              & 0.0137***     &               &               &               &  \\ 
	                                            & (0.0012)      &               &               &               &  \\
{Node strength} \rule{0pt}{3ex}            &               & 0.00679***    &               &               &  \\ 
	                                            &               & (0.00079)     &               &               &  \\ 
{Weighted average} \rule{0pt}{3ex}         &               &               & 0.0147***     &               & 0.0116*** \\ 
{neighbor degree}	                        &               &               & (0.0010)      &               & (0.0011) \\ 
{Eigenvector centrality} \rule{0pt}{3ex}   &               &               &               & 1.073***      & 0.643*** \\ 
{in state}	                                &               &               &               & (0.091)       & (0.099) \\ 
%\emph{Constant} \rule{0pt}{3ex}                 & -0.345***     & -0.365***     & -0.516***     & -0.306***     & -0.446*** \\ 
%	                                            & (0.078)       & (0.078)       & (0.078)       & (0.078)       & (0.078) \\ 
{Fixed effects}  \rule{0pt}{3ex}           & State         & State         & State         & State         & State \\ 
\midrule
{Observations}                             & 6,644         & 6,644         & 6,644         & 6,644         & 6,644 \\ 
$F$-stat                                        & 123.8         & 117.9         & 132.0         & 125.0         & 125.3 \\ 
$R^2$                                           & 0.412         & 0.407         & 0.419         & 0.413         & 0.422 \\ 
{Within} $R^2$                             & 0.169         & 0.163         & 0.179         & 0.171         & 0.184 \\ 
\bottomrule
\end{tabular}
\begin{flushleft}
\footnotesize{Standard errors in parentheses. Significance levels: $^+ p<0.05,\ ^* p<0.01,\ ^{**} p<0.001,\ ^{***} p<0.0001$. \\
Dependent variable in our regressions using individual state data is the inverse hyperbolic sine of COVID cases in the nursing home.}
\end{flushleft}
\label{tab:resultstabledetailed}
\end{table}

% TABLE S2 - BINARY OUTCOME VARIABLE
\begin{table}
\centering \small 
\caption{Covariates of the existence of nursing home COVID-19 cases}
\begin{tabular}{lccccc}
\toprule
& \multicolumn{5}{c}{Dependent variable:  Nursing home has $>0$ cases} \\
\cmidrule(lr){2-6}
                                                & (1)           & (2)           & (3)           & (4)           & (5) \\
\midrule                                                
{Urban indicator}                               & 0.180***      & 0.190***      & 0.164***      & 0.183***      & 0.159*** \\
	                                            & (0.0142)      & (0.0142)      & (0.0141)      & (0.0142)      & (0.0141) \\ 
{Node degree} \rule{0pt}{3ex}                   & 0.00307***    &               &               &               &  \\
	                                            & (0.000320)    &               &               &               &  \\
{Node strength} \rule{0pt}{3ex}                 &               & 0.00135***    &               &               &  \\ 
	                                            &               & (0.000208)    &               &               &  \\ 
{Weighted average} \rule{0pt}{3ex}              &               &               &  0.00399***   &               & 0.00358*** \\ 
{neighbor degree}	                            &               &               & (0.000270)    &               & (0.000297) \\
{Eigenvector centrality} \rule{0pt}{3ex}        &               &               &               & 0.218***      & 0.0860* \\
{in state}	                                    &               &               &               & (0.0240)      & (0.0262) \\ 
%\emph{Constant} \rule{0pt}{3ex}                 & 0.0220        & 0.0173        & -0.0231       & 0.0294        & -0.0138 \\ 
%	                                            & (0.0205)      & (0.0206)      & (0.0205)      & (0.0206)      & (0.0207) \\
{Fixed effects}  \rule{0pt}{3ex}           & State         & State         & State         & State         & State \\ 
{Home demographics} \rule{0pt}{3ex}        & Yes           & Yes           & Yes           & Yes           & Yes \\ 
{CMS quality rating} \rule{0pt}{3ex}       & Yes           & Yes           & Yes           & Yes           & Yes \\ 
\midrule
{Observations}                             & 6,644         & 6,644         & 6,644         & 6,644         & 6,644 \\ 
$F$-stat                                        & 78.46         & 73.36         & 91.23         & 77.51         & 84.65 \\
$R^2$                                           & 0.345         & 0.340         & 0.357         & 0.344         & 0.358 \\ 
{Within} $R^2$                             & 0.114         & 0.107         & 0.130         & 0.113         & 0.132 \\
\bottomrule
\end{tabular}
\begin{flushleft}
\footnotesize{Standard errors in parentheses. Significance levels: $^+ p<0.05,\ ^* p<0.01,\ ^{**} p<0.001,\ ^{***} p<0.0001$. \\
Dependent variable in our linear probability regressions using individual state data is a binary indicator that equals 1 if COVID cases are reported in that nursing home.}
\end{flushleft}
\label{tab:resultstablebinary}
\end{table}

% TABLE S3 - COUNTY FE
\begin{table}
\centering \small 
\caption{Covariates of nursing home COVID-19 cases with county fixed effects}
\begin{tabular}{lccccc}
\toprule
& \multicolumn{5}{c}{Dependent variable:  $sinh^{-1}(Cases)$} \\
\cmidrule(lr){2-6}
                                                & (1)           & (2)           & (3)           & (4)           & (5) \\ 
\midrule
{Node degree}                                   & 0.00385*      &               &               &               &  \\
	                                            & (0.00131)     &               &               &               &  \\
{Node strength} \rule{0pt}{3ex}                 &               & 0.00187$^+$   &               &               &  \\ 
	                                            &               & (0.000843)    &               &               &  \\ 
{Weighted average} \rule{0pt}{3ex}              &               &               & 0.00286$^+$   &               & 0.00167 \\  
{neighbor degree}	                            &               &               & (0.00136)     &               & (0.00142) \\
{Eigenvector centrality} \rule{0pt}{3ex}        &               &               &               & 0.379**       & 0.342* \\ 
{in state}	                                    &               &               &               & (0.110)       & (0.114) \\
%\emph{Constant} \rule{0pt}{3ex}                 & 0.292***      & 0.297***      & 0.254**       & 0.300***      & 0.266** \\ 
%	                                            & (0.0682)      & (0.0683)      & (0.0738)      & (0.0678)      & (0.0739) \\
{Fixed effects}  \rule{0pt}{3ex}                & County        & County        & County        & County        & County \\ 
{Home demographics} \rule{0pt}{3ex}             & Yes           & Yes           & Yes           & Yes           & Yes \\ 
{CMS quality rating} \rule{0pt}{3ex}            & Yes           & Yes           & Yes           & Yes           & Yes \\ 
\midrule
{Observations}                                  & 6,255         & 6,255         & 6,255         & 6,255         & 6,255 \\ 
$F$-stat                                        & 60.90         & 60.50         & 60.44         & 61.28         & 55.84 \\
$R^2$                                           & 0.575         & 0.574         & 0.574         & 0.575         & 0.575 \\ 
{Within} $R^2$                                  & 0.0987        & 0.0981        & 0.0980        & 0.0992        & 0.0993 \\ 
\bottomrule
\end{tabular}
\begin{flushleft}
\footnotesize{Standard errors in parentheses. Significance levels: $^+ p<0.05,\ ^* p<0.01,\ ^{**} p<0.001,\ ^{***} p<0.0001$. \\
Dependent variable is inverse hyperbolic sine of COVID cases in the nursing home using individual state data.}
\end{flushleft}
\label{tab:resultscountyfe}
\end{table}

% TABLE S4 - CMS DATA
\begin{table}
\centering \small 
\caption{Covariates of COVID-19 cases within nursing homes using CMS data}
\begin{tabular}{lccccc}
\toprule
& \multicolumn{5}{c}{Dependent variable: $sinh^{-1}(Cases)$} \\
\cmidrule(lr){2-6}
                                                & (1)           & (2)           & (3)           & (4)           & (5) \\ 
\midrule
{Urban indicator}                               & 0.310***      & 0.349***      & 0.362***      & 0.319***      & 0.309*** \\ 
	                                            & (0.0336)      & (0.0334)      & (0.0333)      & (0.0335)      & (0.0337) \\ 
{Node degree} \rule{0pt}{3ex}                   & 0.0103***     &               &               &               &  \\ 
	                                            & (0.000919)    &               &               &               &  \\
{Node strength} \rule{0pt}{3ex}                 &               & 0.00331***    &               &               &  \\ 
	                                            &               & (0.000463)    &               &               &  \\ 
{Weighted average} \rule{0pt}{3ex}              &               &               & 0.00234***    &               & 0.00137* \\  
{neighbor degree}	                            &               &               & (0.000433)    &               & (0.000443) \\
{Eigenvector centrality} \rule{0pt}{3ex}        &               &               &               & 0.664***      & 0.617*** \\
{in state}	                                    &               &               &               & (0.0634)      & (0.0652) \\ 
%\emph{Constant} \rule{0pt}{3ex}                 & -0.0874       & -0.0892       & -0.112$^+$    & -0.0585       & -0.0765 \\ 
%	                                            & (0.0491)      & (0.0493)      & (0.0496)      & (0.0492)      & (0.0495) \\  
{Fixed effects}  \rule{0pt}{3ex}                & State         & State         & State         & State         & State \\ 
\midrule
{Observations}                                  & 12,775        & 12,775        & 12,775        & 12,775        & 12,775 \\ 
$F$-stat                                        & 152.8         & 145.3         & 143.1         & 151.3         & 139.6 \\  
$R^2$                                           & 0.248         & 0.244         & 0.242         & 0.247         & 0.248 \\  
{Within} $R^2$                                  & 0.116         & 0.111         & 0.109         & 0.115         & 0.116 \\
\bottomrule
\end{tabular}
\begin{flushleft}
\footnotesize{Standard errors in parentheses. Significance levels: $^+ p<0.05,\ ^* p<0.01,\ ^{**} p<0.001,\ ^{***} p<0.0001$. \\
Dependent variable is inverse hyperbolic sine of COVID cases in the nursing home using CMS data.}
\end{flushleft}
\label{tab:resultscms}
\end{table}

% TABLE S5 - NETWORK SUMMARY BY STATE
\begin{table}
\vspace{-0.4cm}
\centering \footnotesize \renewcommand{\arraystretch}{0.9}
\caption{State-level network summary statistics}
\begin{tabular}{lrrrrrrrrrr}
\toprule

      &                &      &        &      &          &      & \multicolumn{2}{c}{Weighted average} & \multicolumn{2}{c}{Eigenvector}      \\
      & \multicolumn{2}{c}{COVID-19 Cases} 
      & \multicolumn{2}{c}{Degree}
      & \multicolumn{2}{c}{Strength} 
      & \multicolumn{2}{c}{neighbor degree}      
      & \multicolumn{2}{c}{Centrality}  \\
\cmidrule(lr){2-3} \cmidrule(lr){4-5} \cmidrule(lr){6-7} \cmidrule(lr){8-9} \cmidrule(lr){10-11}
State & Mean           & SD   & Mean   & SD   & Mean     & SD   & Mean             & SD   & Mean        & SD   \\
\midrule
AL    & 9.1            & 18.9 & 13.3   & 11.1 & 21.3     & 20.5 & 20.8             & 13.1 & 0.19        & 0.25 \\
AR    & 4.3            & 14.7 & 8.3    & 8.5  & 11.5     & 12.7 & 12.2             & 9.7  & 0.15        & 0.25 \\
AZ    & 4.7            & 9.8  & 14.6   & 14.8 & 22.6     & 27.6 & 23.0             & 18.5 & 0.28        & 0.33 \\
CA    & 10.2           & 26.1 & 14.2   & 17.1 & 16.3     & 20.6 & 22.3             & 18.1 & 0.08        & 0.15 \\
CO    & 7.9            & 15.8 & 8.4    & 9.6  & 10.4     & 13.0 & 13.8             & 11.3 & 0.16        & 0.25 \\
CT    & 35.2           & 49.3 & 13.2   & 10.1 & 15.5     & 12.7 & 18.6             & 9.1  & 0.24        & 0.23 \\
DC    & 37.3           & 40.1 & 7.6    & 3.0  & 11.6     & 5.9  & 9.9              & 2.0  & 0.60        & 0.25 \\
DE    & 23.6           & 38.4 & 6.6    & 5.1  & 10.8     & 10.7 & 10.3             & 7.1  & 0.34        & 0.40 \\
FL    & 5.3            & 14.1 & 26.1   & 20.5 & 37.5     & 32.3 & 37.5             & 19.9 & 0.11        & 0.24 \\
GA    & 15.4           & 37.4 & 14.2   & 13.5 & 21.4     & 26.1 & 22.4             & 20.1 & 0.17        & 0.23 \\
IA    & 4.0            & 10.7 & 5.6    & 6.1  & 7.3      & 9.1  & 9.9              & 7.8  & 0.08        & 0.16 \\
ID    & 1.4            & 4.9  & 2.4    & 2.1  & 3.1      & 3.0  & 3.6              & 3.0  & 0.17        & 0.26 \\
IL    & 15.6           & 32.0 & 21.8   & 27.0 & 27.7     & 37.1 & 34.6             & 30.2 & 0.13        & 0.23 \\
IN    & 7.4            & 20.3 & 16.7   & 17.3 & 23.7     & 28.6 & 26.1             & 19.8 & 0.13        & 0.23 \\
KS    & 1.7            & 6.9  & 12.4   & 15.1 & 16.8     & 24.3 & 19.7             & 18.2 & 0.16        & 0.31 \\
KY    & 5.8            & 18.4 & 8.9    & 8.6  & 13.7     & 17.3 & 15.5             & 12.5 & 0.17        & 0.23 \\
LA    & 12.5           & 22.2 & 11.4   & 9.9  & 18.6     & 20.5 & 18.1             & 13.1 & 0.11        & 0.23 \\
MA    & 32.4           & 46.7 & 12.9   & 11.4 & 15.2     & 13.9 & 18.8             & 10.2 & 0.15        & 0.20 \\
MD    & 20.2           & 31.8 & 28.7   & 26.2 & 37.2     & 38.1 & 43.6             & 25.6 & 0.28        & 0.31 \\
ME    & 3.8            & 13.1 & 2.8    & 3.8  & 3.6      & 5.0  & 3.7              & 4.2  & 0.13        & 0.25 \\
MI    & 10.9           & 21.5 & 12.9   & 14.4 & 16.8     & 20.3 & 19.8             & 16.6 & 0.12        & 0.21 \\
MN    & 5.8            & 14.9 & 6.3    & 7.8  & 8.2      & 11.8 & 12.1             & 14.6 & 0.10        & 0.19 \\
MO    & 4.3            & 15.6 & 15.9   & 16.4 & 23.0     & 28.6 & 25.6             & 20.0 & 0.10        & 0.23 \\
MS    & 7.0            & 17.7 & 7.6    & 7.7  & 10.9     & 12.8 & 13.0             & 10.6 & 0.13        & 0.26 \\
MT    & 1.1            & 4.7  & 2.2    & 3.0  & 3.6      & 5.4  & 3.1              & 4.3  & 0.15        & 0.32 \\
NC    & 5.2            & 14.8 & 11.7   & 11.0 & 14.9     & 17.0 & 17.4             & 13.4 & 0.10        & 0.21 \\
ND    & 3.1            & 8.1  & 4.1    & 4.9  & 7.4      & 12.7 & 9.8              & 13.4 & 0.20        & 0.31 \\
NE    & 3.5            & 11.8 & 4.7    & 5.9  & 5.5      & 7.4  & 6.8              & 6.3  & 0.14        & 0.23 \\
NH    & 12.6           & 28.2 & 3.1    & 2.8  & 3.5      & 3.3  & 4.0              & 3.1  & 0.15        & 0.28 \\
NJ    & 36.2           & 50.0 & 26.1   & 19.2 & 35.9     & 30.3 & 38.1             & 18.6 & 0.24        & 0.22 \\
NM    & 3.4            & 12.4 & 3.6    & 3.8  & 5.6      & 7.9  & 4.9              & 4.8  & 0.20        & 0.33 \\
NV    & 21.9           & 38.6 & 6.9    & 6.4  & 10.9     & 11.9 & 10.0             & 7.3  & 0.31        & 0.37 \\
NY    & 25.6           & 46.1 & 19.4   & 20.5 & 28.4     & 39.3 & 41.3             & 36.8 & 0.12        & 0.19 \\
OH    & 5.2            & 15.4 & 18.1   & 15.8 & 23.9     & 24.9 & 29.2             & 17.5 & 0.10        & 0.17 \\
OK    & 2.3            & 8.6  & 11.4   & 10.6 & 15.0     & 15.1 & 17.0             & 10.7 & 0.12        & 0.23 \\
OR    & 3.1            & 7.5  & 4.0    & 4.2  & 4.9      & 5.3  & 5.8              & 4.7  & 0.13        & 0.26 \\
PA    & 16.9           & 32.3 & 14.7   & 14.5 & 20.9     & 29.9 & 25.3             & 24.1 & 0.08        & 0.17 \\
RI    & 22.8           & 38.7 & 12.3   & 10.9 & 15.3     & 15.5 & 21.5             & 12.1 & 0.34        & 0.33 \\
SC    & 8.9            & 23.5 & 9.2    & 7.2  & 14.0     & 12.8 & 13.2             & 8.0  & 0.16        & 0.24 \\
SD    & 1.9            & 6.1  & 1.2    & 1.7  & 1.8      & 3.0  & 2.1              & 2.7  & 0.05        & 0.20 \\
TN    & 2.6            & 8.1  & 13.7   & 10.0 & 21.5     & 20.5 & 22.1             & 12.2 & 0.11        & 0.17 \\
TX    & 3.7            & 16.0 & 20.2   & 19.5 & 29.8     & 68.9 & 53.4             & 93.8 & 0.05        & 0.13 \\
UT    & 1.3            & 3.5  & 5.1    & 5.5  & 7.2      & 8.8  & 8.9              & 6.9  & 0.21        & 0.32 \\
VA    & 9.0            & 23.5 & 12.8   & 10.9 & 17.4     & 16.9 & 18.8             & 10.7 & 0.17        & 0.22 \\
VT    & 8.4            & 18.3 & 0.5    & 1.0  & 0.6      & 1.1  & 0.6              & 1.2  & 0.10        & 0.29 \\
WA    & 6.1            & 18.7 & 6.7    & 6.8  & 8.5      & 9.5  & 9.5              & 7.8  & 0.17        & 0.26 \\
WI    & 3.9            & 10.4 & 6.2    & 9.4  & 8.3      & 13.9 & 10.6             & 12.9 & 0.11        & 0.25 \\
WV    & 3.5            & 11.4 & 6.1    & 6.5  & 10.6     & 15.4 & 13.8             & 11.8 & 0.18        & 0.26 \\
WY    & 1.1            & 3.0  & 0.5    & 0.7  & 0.6      & 0.9  & 0.7              & 0.9  & 0.09        & 0.29 \\
\bottomrule
\end{tabular}
\vspace{-0.2cm}
\begin{flushleft}
\footnotesize{COVID cases include confirmed and suspected cases reported to CMS as of May 31, 2020. \\
\textit{Degree} refers to the number of nursing homes that a particular home is connected to through a smartphone observed in both facilities. \\
\textit{Strength} refers to the total number of smartphones observed in a nursing home and other connected homes. \\
\textit{Weighted average neighbor degree} is the average number of connections a nursing home's neighbor has, weighted by the pair strength. \\
\textit{Eigenvector centrality} measures the extent to which a nursing home's neighbors are highly connected, and is calculated within each state and ranges from 0 to 1.}
\end{flushleft}
\label{tab:statenetwork}
\end{table}

\end{document}